\documentclass[useAMS,usenatbib]{mn2e}
\usepackage{graphics,graphicx,epsfig}
\usepackage[normalem]{ulem}
\usepackage[dvipsnames]{color}
\usepackage{soul,xcolor}
\usepackage{amsmath, amssymb}
\usepackage[]{inputenc,amssymb}
\usepackage{upgreek,textgreek}
\usepackage{booktabs,caption}
\usepackage{amsmath}
\usepackage{multirow}

\setstcolor{red}

\def \be{\begin{equation}}
\def \ee{\end{equation}}
\def \bea{\begin{eqnarray}}
\def \eea{\end{eqnarray}}
\def \etal{{et al.}}

\definecolor{webgreen}{rgb}{0,.5,0}
\definecolor{webbrown}{rgb}{.6,0,0}

\defcitealias{Ewall-Wice2018}{EW18}
\defcitealias{Mirocha2018}{MF18}

\usepackage[%
    colorlinks = true,%
    linkcolor = blue,%
    urlcolor  = blue,%
    citecolor = webgreen,%
    anchorcolor = blue]{hyperref}

\newcommand{\ufhref}[3][blue]{\href{#2}{\color{#1}{#3}}}%

\title[EDGES, radio excess and electron cooling]{Astrophysical radio background 
cannot explain the EDGES 21-cm signal: constraints from cooling of non-thermal electrons}
\author[Sharma]
{Prateek Sharma$^1$\thanks{E-mail: prateek@iisc.ac.in}\\
$^1$Joint Astronomy Programme and Department of Physics, Indian Institute of Science, Bangalore 560012, India}
\begin{document}
\maketitle
\label{firstpage}
\begin{abstract}
Recently the EDGES experiment has claimed the detection of an absorption feature centered at 78 MHz. When interpreted as a signature of 
cosmic dawn, this feature appears at the correct
wavelength (corresponding to a redshift range of $z\approx15-20$) but is larger by at least a factor of two 
in amplitude compared to the standard 21-cm models. One way to explain the  excess radio absorption is by
the enhancement of the diffuse radio background at $\nu = 1.42$ GHz ($\lambda=21$ cm) in the rest frame of the absorbing neutral hydrogen.
Astrophysical scenarios, based on the acceleration of relativistic electrons by accretion on to supermassive 
black holes (SMBHs) and by supernovae (SN) from first stars, have been proposed to produce the enhanced radio 
background via synchrotron emission. In this Letter we show that either
the synchrotron or the inverse-Compton (IC) cooling time for such electrons is at least three orders of magnitude shorter than the 
duration of the EDGES signal centered at $z \approx 17$, irrespective of the magnetic field strength. 
The synchrotron radio emission at 1.42 GHz due to rapidly cooling electrons is $\sim 10^3$ times smaller than the non-cooling estimate.
Thus astrophysical scenarios 
for excess radio background proposed to explain the EDGES signal  
appear very unlikely. 
 \end{abstract}

\section{Introduction} \label{sec:intro}

Very recently the EDGES (Experiment to Detect the Global Epoch of Reionization Signature) experiment has detected a broad 
($\Delta \nu/\nu \approx 1/4$) absorption feature in the residual  
sky brightness temperature centered at $\nu \approx 78$ MHz (\citealt{Bowman2018}). Interpreting this dip in the brightness 
temperature as 21-cm absorption by the diffuse neutral intergalactic medium (IGM) at $z = 
1.42~{\rm GHz}/78~{\rm MHz} -1 \approx 17$ whose spin temperature is coupled to the gas temperature via the Wouthuysen-Field effect
(\citealt{Wouthuysen1952,Field1959}; for a review see \citealt{Pritchard2012}), 
the absorption frequency range is consistent with the standard models for the reionization of the IGM. However, the amplitude of the absorption
feature is at least a factor of two larger than predicted by such models.

The global  
brightness temperature corresponding to the emission/absorption of 21-cm photons for a 
background radiation characterized by a brightness temperature ($T_{\rm bg} = T_{\rm R}+T_{\rm CMB}$; $T_{\rm R}$ and $T_{\rm CMB}$ are the 
brightness temperatures of a diffuse radio background and the CMB) at the redshift of absorption $z$
is given by (Eq. 1 in \citealt{Barkana2018})
\be
\label{eq:T21}
T_{21} = 36 x_{\rm HI} \left( \frac{\Omega_{\rm b} h}{0.0327} \right) \left( \frac{\Omega_{\rm m}}{0.307}\right)^{-1/2} \left(\frac{1+z}{18}\right)^{1/2} \left( 1 - \frac{T_{\rm bg}}{T_{\rm S}}\right)
\ee
in mK, where $x_{\rm HI}$ is the mass fraction of neutral hydrogen, $\Omega_{\rm b}$ and $\Omega_{\rm m}$ are the cosmic mean 
densities of baryons and matter respectively, $h$ is the Hubble parameter in units of 100 km s$^{-1}$ Mpc$^{-1}$, and $T_{\rm S}$ is 
the spin temperature characterizing the level populations of the two hyperfine transition states. The spin 
temperature is expected to lie between the gas kinetic temperature ($T_{\rm K}$) and the background radiation temperature ($T_{\rm bg}$).  

In the standard IGM evolution scenario all the background radiation at the relevant frequencies is due to the CMB. The CMB temperature at 
the relevant  redshift is 
\be
\label{eq:TCMB}
T_{\rm CMB} \approx 49\left(\frac{1+z}{18}\right) {\rm K} 
\ee
and the lowest possible gas kinetic temperature in the standard scenario 
is 7 K (as mentioned in \citealt{Barkana2018}).
 Thus the minimum brightness temperature for the absorption trough at 78 MHz, according to Eq. \ref{eq:T21}, is -216 mK. The 
 brightness temperature measured by EDGES is $-500^{+200}_{-500}$ mK (errors correspond to 99\% [$3\sigma$] confidence intervals; 
 \citealt{Bowman2018}). Thus, even the maximum value of the observationally inferred $T_{21}$ (-300 mK) is lower than the minimum
 according to the standard scenario (-216 mK). From Eq. \ref{eq:T21}, the only way to lower $T_{21}$ is to either raise $T_{\rm bg}$ (e.g., see
\citealt{Ewall-Wice2018,Mirocha2018,Fraser2018,Pospelov2018}) or to lower $T_{\rm S}$ (\citealt{Munoz2018,Barkana2018b,Berlin2018}). In this 
Letter we investigate a subset of the former scenarios.

 The presence of an additional radio background at 21-cm in the rest frame of the absorbing IGM  (characterized by brightness temperature $T_{\rm R}$) 
 will increase the 21-cm absorption signal by an enhancement factor of 
 \be
 \label{eq:E}
 E = \frac{T_{\rm R}/T_{\rm CMB}}{1-T_{\rm S}/T_{\rm CMB}} + 1 \approx \frac{T_{\rm R}}{T_{\rm CMB}} + 1,
 \ee
since $T_{\rm S}/T_{\rm CMB} \approx T_{\rm K}/T_{\rm CMB} \sim 1/7 \ll 1$. Indeed there is an excess radio background measured at
frequencies below 10 GHz, most recently highlighted by the ARCADE 2 experiment (Absolute Radiometer for Cosmology, Astrophysics 
and Diffuse Emission; \citealt{Fixsen2011}; see the recent conference 
summary on this by \citealt{Singal2018}). While this excess radio background cannot be accounted for by extragalactic radio point sources,
most of it may be of Galactic origin (\citealt{Subrahmanyan2013}). The brightness temperature of the excess radio background measured at 78 MHz 
is $\sim 600$ K (see Fig. 1 in \citealt{Singal2018}). We can explain the excess EDGES absorption if only a few K of this (comparable to the CMB 
brightness temperature at $z=0$; see Eq. \ref{eq:E}) is contributed by processes happening earlier than $z\sim17$ (\citealt{Feng2018}).

Astrophysical sources such as accreting supermassive black holes (SMBHs; \citealt{Biermann2014,Ewall-Wice2018}) and supernovae (SN)
from the first stars (\citealt{Mirocha2018}) at $z \gtrsim 17$ can give the required excess radio background. This option, however, requires these
sources to be $\sim 3$ orders of magnitude more efficient radio emitters compared to their low redshift counterparts.

In this Letter we show that the astrophysical mechanisms that require synchrotron emission from relativistic electrons to enhance the radio
background at $z\approx17$ are severely constrained because the cooling time (due to inverse-Compton and synchrotron losses) of these 
electrons is at least three orders of magnitude shorter than the duration of the EDGES absorption trough. 
This implies that the models that do not explicitly account for cooling of non-thermal electrons grossly overestimate the radio synchrotron 
background. 
Although the importance of IC cooling at high redshifts and its contribution to the X-ray background is recognized (e.g., see \citealt{Oh2001,Ghisellini2014}),
here we focus on the cooling argument in the context of the recent EDGES result.

Unless stated otherwise, all quantities are expressed in physical  units in the rest frame of the absorbing gas.

\section{Synchrotron radio background}

In astrophysical scenarios, involving both SMBHs and SN, the excess radio background is produced by incoherent synchrotron emission 
due to relativistic electrons gyrating around magnetic field lines. In this Letter we do not model the radio emissivity due to
accreting SMBHs and star formation at 
$z\sim 17$. In absence of observational constrains, the spectral/redshift variation of emission from these sources is highly uncertain. The radio emissivity models 
are based on extrapolations from low redshifts (e.g.,
see Eqs. 4, 5 in \citealt{Ewall-Wice2018} [hereafter \citetalias{Ewall-Wice2018}]; Eq. 5 in \citealt{Mirocha2018} [hereafter \citetalias{Mirocha2018}]). 
Here we assume that these models for black hole accretion
and star formation at $z \sim 17$, {\em which ignore cooling losses}, can be tuned to raise the radio background at 21-cm. In this Letter the radio source 
evolution is modeled by the 
source term $S_\gamma$ in the one-zone model (Eq. \ref{eq:ngamma}), but we explicitly account for cooling losses which are very important at high redshifts.
We show that the cooling of non-thermal electrons can suppress the radio background by $\sim 10^3$ relative to the non-cooling estimate. 

The synchrotron emission at a frequency $\nu$ is related to the cyclotron 
frequency ($\nu_{\rm cyc} = eB/2\pi m_e c$;  $e$ is the charge of an electron, $B$ is magnetic field strength, $m_e$ 
is electron mass and $c$ is the universal speed of light) by $\nu \sim \gamma^2 \nu_{\rm cyc}$, where $\gamma$ is the Lorentz 
factor of electrons with an isotropic momentum distribution. The Lorentz factor of electrons responsible for synchrotron emission 
at a frequency $\nu=1.42$ GHz is given by
\be
\label{eq:gamma_sync}
\gamma_{\rm syn} \sim 730 \left( \frac{B}{10^{-3} {\rm G}} \right)^{-1/2}.
\ee
The magnetic field strength in our equations refers to the regions in which electrons are  
confined to produce the excess radio background, and not to the strength of a global diffuse magnetic field. The synchrotron photons from
a number of such sources are expected to result in an almost uniform large-scale radio background that may explain the global 21-cm
absorption amplitude.

\subsection{Synchrotron cooling time}

The synchrotron cooling time $t_{\rm syn} \sim \gamma m_ec^2/\gamma^2 u_B \sigma_T c$ ($u_B \equiv B^2/8\pi$ is magnetic energy 
density and $\sigma_T$ is the Thomson scattering cross-section). Expressed in terms of $B$,
\be
\label{eq:tsyn}
t_{\rm syn} \sim 0.05 {\rm Myr} \left( \frac{B}{10^{-3} {\rm G}} \right)^{-3/2}.
\ee

\section{Inverse-Compton considerations}

Typically, the relativistic electrons producing synchrotron emission in radio also emit at much higher frequencies due to the Compton 
upscattering of the background radiation (in our case dominated by the CMB). 

\subsection{IC cooling time}

The IC cooling time for relativistic electrons is $t_{\rm IC} = \gamma m_e c^2/ \gamma^2 u_{\rm bg} \sigma_T c$ ($u_{\rm bg}$ is the energy
density of the background photons). The ratio of the IC cooling time to the synchrotron cooling time {\em for the 
electrons producing  the excess radio background} is 
\be
\label{eq:IC_sync_ratio}
\frac{t_{\rm IC}}{t_{\rm syn}} \sim \frac{u_{\rm B}}{u_{\rm bg}} \sim 0.91 \left( \frac{B}{10^{-3} {\rm G}} \right)^{2} \left(\frac{1+z}{18}\right)^{-4},
\ee
where we have assumed $u_{\rm bg} = u_{\rm CMB} = aT_{\rm CMB}^4$ ($a$ is the radiation constant). Thus the IC cooling time for these
electrons (combining Eq. \ref{eq:IC_sync_ratio} with Eq. \ref{eq:tsyn}) is 
\be
\label{eq:tIC_syn}
t_{\rm IC} \sim 0.046 {\rm Myr} \left( \frac{B}{10^{-3}{\rm G}} \right)^{1/2} \left(\frac{1+z}{18}\right)^{-4}.
\ee

\begin{figure}
\centering
\includegraphics[height= 2.5in,width=3.4in]{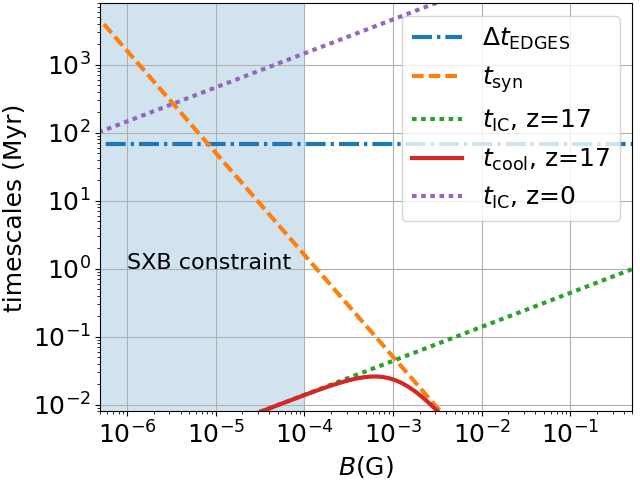}
\caption{Important timescales for the relativistic electrons producing the excess radio synchrotron background 
as a function of the magnetic field strength: the duration of the EDGES signal ($\Delta t_{\rm EDGES}$; Eq. \ref{eq:dt_EDGES}),
the synchrotron cooling time ($t_{\rm syn}$; Eq. \ref{eq:tsyn}), the IC cooling time ($t_{\rm IC}$ at $z=0,~17$; Eq. \ref{eq:tIC_syn}), and 
the cooling time $t_{\rm cool}$ (Eq. \ref{eq:tcool}) at $z=17$. The relativistic electrons will cool at the shorter of the synchrotron and IC cooling timescales, 
which at $z=17$ is at least $\sim10^3$ shorter than the duration of the EDGES absorption for all $B$. 
Magnetic field strength lower than $\sim 10^{-4}$ G (indicated by 
the blue shaded region) is ruled out from the soft X-ray background constraint (see section \ref{sec:SXB}). An important timescale, the electron replenishment
timescale in AGN/SN, is highly uncertain and not shown here but discussed in section \ref{sec:one-zone}.
}
\label{fig:timescales}
\end{figure}

Figure \ref{fig:timescales} shows the synchrotron and IC cooling timescales as a function of the magnetic field strength at $z=17$ and $z=0$.
Note that  at $z=0$ the IC and synchrotron cooling times cross at the field strength of $B \approx 3~\mu$G. The same crossover 
at $z=17$ occurs at $B \sim 10^{-3}$ G. Also note that the maximum value of the cooling time 
($t_{\rm cool} \approx {\rm min}[t_{\rm syn}, t_{\rm IC}]$; see Eq. \ref{eq:tcool}) 
at $z=17$ is about four orders of magnitude shorter than at $z=0$, implying that the cooling losses are much more important at higher redshifts than now.

\subsection{Soft X-ray background}
\label{sec:SXB}
Let us, for now,  
assume that synchrotron radio emission indeed produces the requisite radio background at $z \sim 17$. The
same electrons are expected to upscatter the CMB photons and produce a uniform background at a frequency (see Eq. \ref{eq:gamma_sync})
\be
\label{eq:IC_R}
\nu_{\rm IC} \sim \gamma_{\rm syn}^2 \nu_{\rm bg} \sim 1.5\times 10^{18} {\rm Hz}   \left( \frac{B}{10^{-3} {\rm G}} \right)^{-1} \left(\frac{1+z}{18}\right),
\ee
which corresponds to 6.4 keV for the fiducial parameters. This IC background will be redshifted to soft X-rays ($\sim 0.36$ keV, or equivalently, $\nu \sim 8 
\times 10^{16}$ Hz) at $z=0$.

The synchrotron emissivity of relativistic electrons is given by
\be
\label{eq:syn_em}
\epsilon_{\nu, \rm syn} \sim \gamma_{\rm syn}^2 u_{\rm B} \sigma_T c \left[ \frac{dn}{d\gamma} \frac{d\gamma}{d\nu} \right]_{\nu = \gamma_{\rm syn}^2 \nu_{\rm cyc}},
\ee
where $dn/d\gamma \propto \gamma^{-p}$ is a power-law distribution of the relativistic electrons. The IC emissivity produced by the upscattering 
of the CMB {\em by the same electrons} is related to it by
$\epsilon_{\nu, \rm IC}/\epsilon_{\nu, \rm syn}  \sim u_{\rm CMB} \nu_{\rm cyc}/u_{\rm B} \nu_{\rm CMB}$ (a consequence of $\nu_{\rm syn/IC} \sim 
\gamma^2 \nu_{\rm cyc/CMB}$). Therefore the ratio of IC and synchrotron 
emissivities per logarithmic interval in frequency is given by
\be
\label{eq:IC_em}
\frac{(\nu \epsilon_\nu)_{\rm IC}}{(\nu \epsilon_\nu)_{\rm syn}} = \frac{u_{\rm CMB}}{u_{\rm B}} \sim 1.1 \left( \frac{B}{10^{-3} {\rm G}} \right)^{-2}  \left(\frac{1+z}{18}\right)^4. 
\ee
Now if this IC emission in X-rays travels to $z=0$ without getting absorbed, for our fiducial $B$ we expect a soft X-ray background (SXB) 
with $\nu F_\nu$ comparable
to the radio background at $\nu \sim 78$ MHz measured at $z=0$. Thus the spectral energy surface brightness density ($\nu I_\nu$) at $\sim 8 \times10^{16}$ Hz
(0.36 keV) contributed by the synchrotron radio emitting electrons due to IC upscattering of the CMB is $\sim 3 \times 10^{-4}$ nW m$^{-2}$ Sr$^{-1}$, 
a few \% of the observed SXB (e.g., see Fig. 2 in \citealt{Singal2018} and compare the backgrounds at  $10^8$ Hz and $10^{17}$  Hz). The constraint of
not overproducing the SXB implies that $B \gtrsim 10^{-4}$ G in the sources responsible for synchrotron radio emission (the blue shaded region in Fig. 
\ref{fig:timescales} marks the field strengths ruled out by this constraint).   

Moreover, such an SXB will have implications on the X-ray heating of the IGM, the evolution of the spin temperature, and the appearance of 
the 21-cm feature (see Eq. \ref{eq:T21}). We do not explore these issues in this Letter (for a discussion, see \citetalias{Ewall-Wice2018,Mirocha2018}).

\section{Cooling constraint}

\begin{figure}
\centering
\includegraphics[height= 2.5in,width=3.4in]{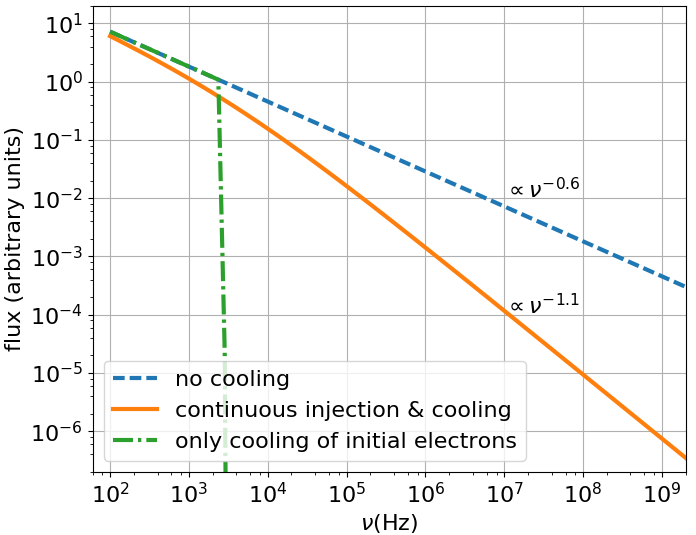}
\caption{The average spectral energy distribution (SED) of the radio synchrotron sources at 18.25 Myr ($< \Delta t_{\rm EDGES}$; corresponding to the 
cooling time of $\gamma=1$ electrons) under different scenarios of electron cooling: (i) no cooling; (ii) continuous injection of electrons with cooling; and (iii) 
only cooling (and no continuous injection) of the initially 
relativistic electrons. For (ii) the spectral index ($d\ln F_\nu/d\ln \nu$) becomes steeper by 0.5 beyond the cooling break; for (iii) there is no emission at frequencies
higher than the cooling cut-off. Since the minimum value of $\Delta t_{\rm EDGES}/t_{\rm cool} > 10^3$ for all field strengths (see Fig. \ref{fig:timescales}), the 
cooling break at $\sim \Delta t_{\rm EDGES}$ occurs at a very low frequency, and the flux at 1.42 GHz for (ii) is $\sim 10^3$ lower than (i).
}
\label{fig:Fnu}
\end{figure}

The age of the Universe at the redshift of the EDGES absorption trough ($z \sim 17$), assuming a flat matter-only Universe (a good 
assumption at those redshifts), is
\be
\label{eq:tage}
t_{\rm age} \sim 180 {\rm Myr} \left( \frac{1+z}{18} \right)^{-3/2}.
\ee
The EDGES absorption trough is centered at $z \approx 17$ and has a duration $\Delta z/z \approx 1/4$. For a flat matter-only Universe
$\Delta t/t = -3 \Delta z/2(1+z)$, and therefore the time duration of the EDGES signal is
\be
\label{eq:dt_EDGES}
\Delta t_{\rm EDGES} = \frac{3}{2} t_{\rm age} \frac{\Delta z}{1+z} \approx 67.5~{\rm Myr} \left( \frac{1+z}{18} \right)^{-3/2} \frac{\Delta z/z}{1/4}.
\ee

Now if an enhanced radio background has to explain the EDGES absorption amplitude, it must be present for at least the duration of $\Delta t_{\rm EDGES}$.
If the relativistic electrons are accelerated only at the beginning of the absorption feature at $z \approx 20$ but are not replenished,
they will cool off in $\lesssim 0.025$ Myr irrespective of $B$ (see Fig. \ref{fig:timescales}) and therefore the required radio background cannot 
be sustained for $\Delta t_{\rm EDGES}$.  This implies that we need 
continuous injection of relativistic electrons to replenish  cooling losses. But even with the continuous injection of electrons, cooling is expected to steepen 
the radio spectrum at the relevant frequencies, as we show next.

\subsection{A one-zone model}
\label{sec:one-zone}

The standard one-zone model for the evolution of $n_\gamma \equiv dn/d\gamma$ is given by (e.g., see Eq. 14 in \citealt{Oh2001}; 
see also \citealt{Sarazin1999})
\be
\label{eq:ngamma}
\frac{\partial n_\gamma}{\partial t} + \frac{\partial}{\partial \gamma} (\dot{\gamma} n_\gamma) = S_\gamma - \frac{n_\gamma}{t_{\rm esc}},
\ee
where $\dot{\gamma} = -\gamma/t_{\rm cool}$, 
\be
\label{eq:tcool}
t_{\rm cool} = \frac{1}{t_{\rm syn}^{-1} + t_{\rm IC}^{-1} }
\ee
is the cooling time due to synchrotron and IC cooling (we ignore adiabatic cooling), $S_\gamma$ is the source term for relativistic electrons, 
and $t_{\rm esc}$ is the escape timescale of relativistic electrons. 
The spectral index $\alpha_\nu \equiv d\ln F_\nu/d\ln \nu$ and the electron power-law index ($p$; $n_\gamma \propto \gamma^{-p}$) are 
related as $\alpha_\nu = -(p-1)/2$ (e.g., see \citealt{Rybicki1986}).

If there is no injection (i.e., $S_\gamma=0$) but only cooling of an initial population of relativistic electrons, a cooling
cut-off occurs at $\gamma$ (the corresponding $\nu = \gamma^2 \nu_{\rm cyc}$) for which the cooling time equals the age of electrons; there are no electrons 
with a higher $\gamma$. The SED for this case is shown by the green dot-dashed line in Fig. \ref{fig:Fnu}. 

{\em Steady solution with injection and no cooling:} In absence of cooling, the source term needed to produce an electron number density sufficient to produce the 
required radio background ($n_\gamma \propto \epsilon_{\nu, \rm syn}$; see Eq. \ref{eq:syn_em}) is given by $S_\gamma = n_\gamma/t_{\rm esc}$. Thus, in steady 
state the electron replenishment time equals the escape time ($t_{\rm esc}$). 
Extrapolating from $z \sim 0$, \citetalias{Ewall-Wice2018} and \citetalias{Mirocha2018} use a shallow spectral index ($\alpha_\nu=-0.6,~-0.7$ respectively) for 
the radio emission from their $z \approx 17$
sources. The corresponding electron indices are $p=$2.2, 2.4, consistent with diffusive shock acceleration. These works do not account for cooling losses 
of the relativistic electrons, which are clearly very important at $z \sim 17$ as compared to $z \sim 0$ (see Fig. \ref{fig:timescales}). 
Now we consider the effects of cooling.

{\em Solution with injection and cooling:} In presence of cooling, a cooling break appears for the solution of Eq. \ref{eq:ngamma} at 
a $\gamma$ for which the cooling time 
equals the age of electrons ($t$). For $\gamma$s smaller than the cooling break $n_\gamma = S_\gamma t_{\rm esc}$, the same as in the no-cooling case.
However, for $\gamma$s larger than the cooling break $\partial/\partial \gamma (\dot{\gamma} n_\gamma) \approx S_\gamma$. Since $S_\gamma$ is assumed 
to be a power-law (with $p=2.2$) and $\dot{\gamma} \propto -\gamma^2$ (true for both synchrotron and IC losses), the slope of $n_\gamma$ steepens by 
unity beyond the cooling break and the slope of the SED steepens by 0.5.

Figure \ref{fig:Fnu} shows the radio SEDs for models based on Eq. \ref{eq:ngamma} (synchrotron emissivity and $n_\gamma$ are related by Eq. \ref{eq:syn_em})
without injection (green dot-dashed line), with injection but no cooling (blue dashed line), and with injection 
and cooling (orange solid line). 
The dashed line shows the SED assumed by \citetalias{Ewall-Wice2018} with a large enough $S_\gamma$ ($S_\gamma \propto \gamma^{-2.2}$, equivalently 
$S_\nu \propto \nu^{-0.6}$), corresponding to their PopIII scenario (see their Fig. 1) that 
produces sufficient radio background. The solid line shows the SED with the same $S_\gamma$ but affected by cooling 
(assuming $B=10^{-3}$ G corresponding 
to the longest  $t_{\rm cool} \approx t_{\rm syn}/2 \approx t_{\rm IC}/2 \approx 0.025$ Myr; see Fig. \ref{fig:timescales})
after a duration of $0.025 \times 730 \approx 18.25$ Myr ($< \Delta t_{\rm EDGES}=67.5$ Myr; Eq. \ref{eq:dt_EDGES}), 
the time at which the cooling break for electrons reaches $\gamma=1$ ($\nu = \nu_{\rm cyc} \approx 2500$ Hz).\footnote{Synchrotron self absorption will become 
important at such low frequencies but the SED at $1.42$ GHz (in the optically thin regime) will be unaffected.}

The cooling time for non-relativistic electrons emitting cyclotron photons becomes independent of 
energy and the $t_{\rm cool} \propto \gamma^{-1}$ scaling breaks down, but the SED at $\sim \Delta t_{\rm EDGES}$ will definitely be below the orange solid line in 
Figure \ref{fig:Fnu}. Thus reading off the values at 1.42 GHz for the orange and blue lines in Figure \ref{fig:Fnu}, we conclude that ignoring cooling losses 
overestimates the radio synchrotron flux by $\gtrsim 10^3$. Therefore, the source radio emissivities (equivalently $S_\gamma$) need to be boosted by $\sim 10^3$ 
to reproduce the EDGES absorption in presence of realistic cooling.
This essentially rules out synchrotron
radio background as a solution to the enhanced 21-cm absorption seen by EDGES because the models are already fine tuned to produce $\sim 10^3$ larger 
radio emission compared to $z\sim 0$ observations.

\section{Conclusions}
We conclude that astrophysical particle accelerators (first stars and supermassive black holes), with reasonable extrapolation from $z \sim 0$, 
cannot produce the radio synchrotron
background at 1.42 GHz comparable to the CMB brightness temperature. Such a radio background is invoked by some models 
(e.g., \citetalias{Ewall-Wice2018,Mirocha2018}) to explain the excess 21-cm absorption signature claimed by the EDGES experiment.
The principal difficulty is that the cooling time (shorter of the synchrotron and inverse-Compton cooling times) of the relevant relativistic electrons 
is at least three orders of magnitude shorter than the duration of the EDGES signal (see Fig. \ref{fig:timescales}).
To get the required radio background, various astrophysical scenarios have to enhance the radio emissivity 
by $\sim 10^3$ compared to the $z\sim0$ models. In the 
presence of non-thermal cooling losses considered in this Letter the required enhancement is expected to be $\gtrsim 10^6$, 
which seems almost impossible. Of course, there is the additional constraint from the soft X-ray background (see section \ref{sec:SXB}).

We note that the constraints in this Letter do not apply to scenarios in which the excess radio background is not produced by
relativistic electrons (e.g., \citealt{Fraser2018,Pospelov2018}). 
With stringent constraints on dark matter cooling (\citealt{Munoz2018,Barkana2018b,Berlin2018}) 
and on astrophysical models based on excess radio background, it is imperative that the EDGES signal be confirmed
with other experiments. Thankfully there are several such ongoing experiments (e.g., see \citealt{Bernardi2016,Voytek2014,Singh2017}).

We end by noting that the arguments in this Letter can be used to put tight constraints on the background radiation at high redshifts, 
independent of the fate of the EDGES signal.

\section*{Acknowledgements}
We thank  Siddhartha Gupta and Biman Nath for helpful discussions and encouragement. We also thank Rohini Godbole and Nirupam 
Roy for organizing a discussion on the EDGES result. We are grateful to Aaron Ewall-Wice and an anonymous referee for useful comments. 
We acknowledge an India-Israel joint research grant (6-10/2014[IC]).

\end{document}